\documentclass[12pt]{article}
\usepackage{amsmath,amssymb,epsfig}
\usepackage{graphicx,floatflt,subfigure}
\usepackage{epstopdf}
\usepackage{color}
\input{colordvi.tex}
\usepackage[table]{xcolor}

\definecolor{shadecolor}{gray}{0.80}

\def\nonu{\nonumber}

\usepackage[table]{xcolor}
\usepackage{amsmath,amssymb,epsfig}
\usepackage{framed}
\usepackage{color}
\definecolor{shadecolor}{gray}{0.80}
\usepackage{fancybox,ascmac}
\usepackage[all]{xy}
\usepackage{colortbl}

\usepackage{amsmath,caption,booktabs}
\usepackage[table]{xcolor}

\makeatletter

\renewcommand\section{\@startsection {section}{1}{\z@}%
                                   {-3.5ex \@plus -1ex \@minus -.2ex}%
                                   {2.3ex \@plus.2ex}%
                                   {\normalfont\large\bfseries}}

\renewcommand\subsection{\@startsection{subsection}{2}{\z@}%
                                     {-3.25ex\@plus -1ex \@minus -.2ex}%
                                     {1.5ex \@plus .2ex}%
                                     {\normalfont\normalsize\bfseries}}

\makeatother

\begin{document}

\baselineskip=18pt  
\numberwithin{equation}{section}  
\allowdisplaybreaks  



%
%


\thispagestyle{empty}

\vspace*{-2cm}
\begin{flushright}
\end{flushright}

\begin{flushright}
\end{flushright}

\begin{center}

\vspace{2cm}

{\bf\Large Catalytic Creation of Bubble Universe }
\vspace*{0.2cm}

{\bf\Large  Induced by Quintessence in Five Dimensions }

\vspace{1.3cm}

{\bf
Issei Koga$^{1}$ and Yutaka Ookouchi$^{2,1}$} \\
\vspace*{0.5cm}

${ }^{1}${\it Department of Physics, Kyushu University, Fukuoka 810-8581, Japan  }\\
${ }^{2}${\it Faculty of Arts and Science, Kyushu University, Fukuoka 819-0395, Japan  }\\

\vspace*{0.5cm}

\vspace*{0.5cm}

\end{center}

\vspace{1cm} \centerline{\bf Abstract} \vspace*{0.5cm}

We investigate the bubble nucleation in five dimensional spacetime catalyzed by quintessence. We especially focus on decay of a metastable Minkowski vacuum to an anti-de Sitter vacuum and study dynamics of the bubble on which four dimensional expanding universe is realized. We also discuss the trans-Planckian censorship conjecture and impose a constraint on the parameter space of the catalysis. As an application of this model, we propose an inflation mechanism and an origin of the dark energy in the context of quintessence in five dimensions. 

\newpage
\setcounter{page}{1} 



\section{Introduction}

The structure of vacua in unified theories has attracted wide attention recently. Especially, after the discovery of the Higgs particle and precise measurements of top quark mass, it have been believed that our universe is metastable even within the standard model \cite{MetaUniverse}. This possibility was firstly pointed out by \cite{FirstPoint,Coleman,CDL}.  Related works on this instability of our universe has been done from various points of view \cite{HiggsDecay}. In string theories, the vacuum structure becomes more involved, which is known as the string landscape (see \cite{StringLandscape} for example), and has been discussed recently in the context of swampland conjectures \cite{Swampland}. One of the remarkable conjectures is the de Sitter conjecture \cite{NoDeSitter} which prohibits making a four dimensional de Sitter space by compactifying the internal space.  See \cite{SwamplandReview} for reviews and references therein. This conjecture is controversial and still under debate. However, it would be interesting to explore another realizations of our universe in string theories in light of the conjectures. An interesting avenue was open up by the authors of \cite{Dan}, in which they realized four dimensional universe on a bubble in five dimensions created by a decay process of metastable anti-de Sitter (AdS) vacuum. The radiation and matter in four dimensions are realized in terms of a black hole and a string cloud in five dimensions. Naively, since the bubble is the boundary of two AdS spaces, the four dimensional gravity can be localized on the bubble in the same spirit as Randall-Sundrum \cite{RS} scenario (see \cite{RSnew,NewDan} for more recent studies on this issue). The catalytic effects caused by the string cloud and the black hole in this context was recently discussed in \cite{KO1} and showed that the catalysis provides a kind of the selection rule to the cosmological constant on the bubble universe. This paper can be regarded as a continuation of this study and we try to engineer inflation sector and the dark energy in this context.

Catalytic effects in field theories were firstly pointed out by \cite{Stein} and discussed in various contexts such as realistic model building \cite{CatalysisPheno} and decay processes in stringy theories \cite{CatalysisString}. Also, this idea have been discussed in the context of gravitational theories \cite{Gregory14} initiated by \cite{Hiscock}.  In this paper, we study vacuum decay along the lines of these papers, especially, by using the method developed in \cite{Gregory14} to treat a singular bounce solution. Recently, catalytic effects in gravitational theories have been discussing in various contexts  \cite{Oshita,Gregory15,GregoryHiggs1,GregoryHiggs2,Gregory19,KO2}.  

In this paper, we introduce quintessence in this scenario and discuss catalysis induced by it. A discontinuity of quintessence on the bubble can be interpreted as a four dimensional quintessence. One of the remarkable features of the quintessence is time dependence of $w_{(4)}$ in the state equation. Even if $w_{(4)}\simeq -1$ at the present age of the universe, it cloud be larger at the early stage. It would be interesting if quintessence can play a role of catalyst when the bubble universe is created. Moreover, we will use the quintessence to engineer the inflation at the early stage and the dark energy at the late stage of the universe. 

The organization of this paper is as follows: In section 2, we review black hole solutions spherically surrounded by quintessence in four and five dimensions. Also we briefly review junction conditions for connecting two solutions with different parameters. Then, we show how to compute the bounce action for the decay of metastable vacuum and discuss a recent development on the bounce action for a solution with singularities  along the lines of \cite{Coleman,Gregory14,Gregory15}. In section 3, we show catalytic effects induced by quintessence for the decay process of a metastable Minkowski vacuum to an anti-de Sitter (AdS) vacuum. In section 4, we consider a model including two types of quintessence. First, we study catalytic effects induced by the quintessence and show a selection rule of bubbles in five dimensions. Then, we further impose a constraint by using the trans-Planckian censorship conjecture \cite{TCC}. After that, we discuss an application of this model to a realization of inflation and the dark energy on the bubble universe. We use the freezing-type quintessence as an inflaton and thawing-type as the dark energy at the present age. The section 5 is devoted to conclusions and summary. In appendix A, we quickly review the Coleman-de Luccia (CDL) bounce action in five dimensions \cite{CDL,KO1}.

\section{General arguments}

In this paper, we incorporate inflation and the dark energy with the bubble universe realized in five dimensions \cite{Dan}. Toward this goal, we use quintessence as candidates for inflaton and the dark energy. We treat spherically symmetric gravitational solutions for quintessence and study junctions of two solutions with different parameters. Hence, we first review the solutions in four and five dimensions, then, show basic formulae which will be used in computing the bounce action in the next section.  To compute the bounce action, by using the method developed by Coleman \cite{Coleman}, we solve the equation of motion for the bubble, which is the junction surface separating two regions, in Euclideanized theory and plug the solution back into the action. We also comment that the singularity at the origin of the solutions does not contribute to the bounce action.   

\subsection{Gravitational solution for quintessence}

Here, we quickly review the solutions for quintessence in four and five dimensions along the lines of \cite{Kiselev,Higher}. First, we treat four dimensions. In the standard cosmology, the state equation relates the pressure $p$ with the energy density $\rho$; $p=w_{(4)}\rho$. To distinguish from five dimensional quintessence, we add the subscript. In four dimensions, the cosmological constant, radiation and matter correspond to $w_{(4)}=-1,1/3,0$ respectively. We use the terminology ``quintessence'' in a broad sense in which all the states except these three cases are quintessence. Moreover, quintessence states can be divided into two parts by the acceleration of the universe. From the Friedmann equation,  
\begin{eqnarray}
{\ddot{a} \over a}=-{4\over 3} G_4 (1+3w_{(4)})\rho  \quad  \quad  \mbox{(in\ four\ dimensions)}~,
\end{eqnarray}
we see that when $w_{(4)}\le -1/3$, the universe is accelerated by quintessence while $w_{(4)}> -1/3$ corresponds to deceleration.

\begin{figure}[t]
\begin{center}
 \includegraphics[width=.4\linewidth]{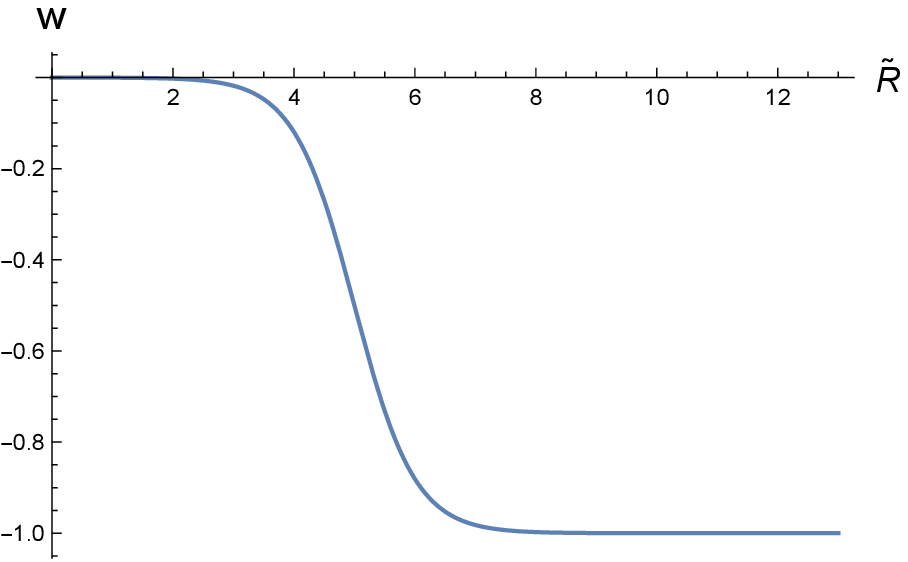}\hspace{1cm}
\includegraphics[width=.4\linewidth]{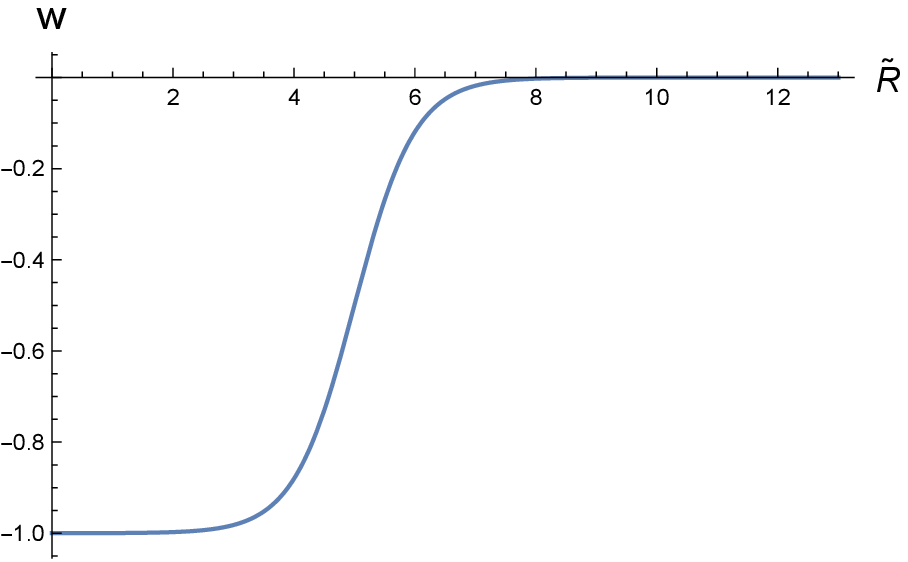}
\vspace{-.1cm}
\caption{\sl Schematic picture of $w$. We assume that $w$ in five dimensions varies as a function of scale factor. In the freezing model, it starts around $0$ and goes to $-1$, while in the thawing, it starts from $-1$ and goes to $0$. In section 3 and 4, we discuss four dimensional universe realized on the bubble created in five dimensions. In this case, $\widetilde{R}$, which is normalized scale factor \eqref{normalization}, corresponds to the size of the bubble in five dimensions. }
\label{FigOurAssumption}
\end{center}
\end{figure}

Time dependence of $w_{(4)}$ varies from model to model and various kinds of phenomenological models for quintessence has been proposed. See \cite{QuinReview} for reviews. Among them, the freezing and thawing models are suitable for our purpose. So we discuss them by idealizing the dependence of $w_{(4)}$ as functions shown in figure \ref{FigOurAssumption} for the sake of simplicity. In the freezing model, $w_{(4)}$ starts around zero. It gradually gets smaller and eventually reaches $-1$. In the thawing model, it starts around $-1$ and finally becomes zero. Note that the figure \ref{FigOurAssumption} is just a schematic picture to demonstrate our assumption, so the numbers in the figure do not have any sense. Throughout this paper, we simply assume the scale factor dependence of $w_{(4)}$ as in figure \ref{FigOurAssumption} without specifying explicit models.  

Black hole solutions surrounded by quintessence was shown in \cite{Kiselev}
\begin{eqnarray}
ds^2=-f_K(r) dt^2 +f_K^{-1}(r) dr^2 + r^2 d\Omega_2^2~,\label{4Dsol}
\end{eqnarray}
where $d\Omega_2^2$ is the two dimensional round metric and
\begin{eqnarray}
f_K(r) =1-{r_g^2\over r^2}-\sum_n \Big( {r_n \over r}\Big)^{3w_{(4)}^{(n)} +1}~.
\end{eqnarray}
The label $n$ runs from $1$ to the total number of quintessence. Although the solutions corresponding to accelerated universe have both a cosmological and an event horizons,  the solutions for $-{1/ 3} < w_{(4)} < 0$, there is no cosmological horizon. 

These quintessential solutions were extended to higher dimensions \cite{Higher}. Here, let us focus on five dimensions for our purpose. In this case, the solution is given by  
\begin{equation}
ds^2= -f(r) dt^2 +{dr^2 \over f(r)}+r^2d\Omega_3^2\, , \label{5Dsol}
\end{equation}
where $d\Omega_3^2$ is the three dimensional round metric and the function $f(r)$ is 
\begin{eqnarray}
f(r)&=&1-{r_{BH}^2\over r^2}-{\Lambda^{(5)}\over 6}  r^2-\sum_n{q^{(n)}\over r^{4w^{(n)}+2}}\, .
\end{eqnarray}
Here, we wrote the contribution of the cosmological constant separately, because we will deal with models having both cosmological constant and quintessence later. In five dimensions, the accelerated universe corresponds to $w<-{1/ 2}$. 

In the solutions \eqref{4Dsol} and \eqref{5Dsol}, contributions from black holes exist. These are important in making spherically symmetric quintessence in a realistic situation. However, in the discussions below, we put $r_{BH}=0$ for the sake of simplicity and illustrate the catalytic effect induced by quintessence.  As we will see in section 3 and 4, inhomogeneity of these solutions, which can be seen in the singularity at the origin,  enhances the decay rate of the metastable vacuum.

\subsection{Junction condition and equation of motion for bubble}

In discussing the bubble nucleation, we study a junction surface of two different spherically symmetric solutions. Over the surface, there are discontinuities of physical quantities such as the curvature and the energy-momentum tensor. The discontinuity of the curvature can be  expressed in terms of the extrinsic curvature $K_{ij}$ as follows: The Einstein equation on the surface can be reduced to
\begin{eqnarray}
  K^+_{ij}-K^-_{ij}=8\pi G_5\left(S_{ij}-\frac{1}{3}\gamma_{ij}S\right)\, ,
\end{eqnarray}
where $i,j$ runs from 1 to the dimension of the bubble. This is known as the Israel's junction condition \cite{Israel}. Hereafter, we write the subscript $+\, (-)$ for quantities outside (inside) the bubble. For the sake of simplicity, we impose thin-wall approximation and use $S_{ij}=-\sigma \gamma_{ij}$. With this notation and by taking the trace of the extrinsic curvature, the junction condition becomes 
\begin{eqnarray}
  K^+-K^-={32\pi G_5 \over 3} \sigma \, ,\label{IScond}
\end{eqnarray}
where we used $S=-4\sigma$. 

For later convenience, let us modify the expression and make the condition a bit simpler form. We basically adopt the same notations used in \cite{Gregory15} and define $\eta=\bar{\sigma}l$, $\bar{\sigma}={4\pi G_5 \sigma /3}$ and   
\begin{equation}
l^2={6 \over \Delta \Lambda^{(5)}} \ ,\quad \gamma= {4\bar{\sigma}l^2 \over 1+4\bar{\sigma}^2 l^2}\ ,\quad \alpha^2 =1+{\Lambda^{(5)}_- \gamma^2 \over 6}\, ,
\end{equation}
where $\Delta \Lambda^{(5)} =\Lambda^{(5)}_+-\Lambda^{(5)}_-$. Two geometries with different parameters are connected at $r=R(\lambda)$ ($\lambda$ is the proper time on the bubble), thus the induced metric on the bubble becomes the Friedmann type form, 
\begin{equation}
ds^2=-d\lambda^2 +R^2(\lambda) d\Omega^2_3\, .
\end{equation}

To estimate the decay rate for this bubble nucleation, by following the Coleman's method \cite{Coleman}, we introduce the Euclidean-time defined by $t=-i\tau$ and look for a classical solution in the Euclideanized theory. By computing the extrinsic curvature with the notations above, \eqref{IScond} can be written as
\begin{eqnarray}
  \frac{1}{ R}(f_+\dot \tau_+-f_-\dot \tau_-)=-\frac{8\pi G_5}{3}\sigma\, ,
\end{eqnarray}
where $\dot{\tau}_\pm$ satisfy the following relations
\begin{eqnarray}
  f_\pm \dot{\tau}_\pm^2+{\dot{R}^2 \over f_\pm}=1\, .
\end{eqnarray}
Using these expressions and introducing $\bar f=(f_++f_-)/2$ and $\Delta f=f_+-f_-$, we obtain the equation of motion for the bubble 
\begin{equation}
\dot{R}^2 = -\bar{\sigma}^2 R^2 +\bar{f}- {(\Delta f)^2 \over  16   \bar{\sigma}^2 R^2}\, . 
\end{equation}
For numerical calculations, we define dimensionless coordinates as in \cite{Gregory15},
\begin{equation}
\widetilde{R}={ \alpha R \over \gamma}\  ,\quad\  \quad \tilde{\lambda}={\alpha \lambda \over \gamma}\  ,\quad \  \quad \tilde{ \tau}={\alpha \tau  \over \gamma} \, . \label{normalization}
\end{equation}
It is also convenient to introduce dimensionless parameters for quintessence,
\begin{eqnarray}
Q_{\pm(n)}^{(w)}=\Big({\alpha \over \gamma} \Big)^{4w+2}q_{\pm(n)}~,
\end{eqnarray}
and to express the functions $f_\pm$ in terms of them
\begin{eqnarray}
f_\pm(\tilde{R})=1 +\Big({\gamma \over l_\pm \alpha} \Big)^2 \tilde{R}^2 -\sum_n{Q^{(w)}_{\pm (n)} \over \tilde{R}^{4w^{(n)}+2}} \ .
\end{eqnarray}
Throughout this paper, since we assume vacuum decay between AdS (or Minkowski) spacetimes in five dimensions, we define the typical length of the spacetimes as 
\begin{eqnarray}
l_\pm^2 =-{6\over \Lambda^{(5)}_\pm}~.
\end{eqnarray}
The equation of motion for the radius of the bubble can be expressed as
\begin{eqnarray}
(\dot{\widetilde{R}})^2=1-\widetilde{R}^2-\sum_n \Big(\bar{Q}_{(n)}+{\Delta Q_{(n)}\over 8\eta^2} \Big){1\over \widetilde{R}^{4w^{(n)}+2}}-{1\over 16\eta^2}\Big({l\alpha \over \gamma} \Big)^2 \Big( \sum_n {\Delta Q_{(n)} \over \widetilde{R}^{4w^{(n)}+3} }\Big)^2~, \label{GeneralEQ}
\end{eqnarray}
where we introduced
\begin{eqnarray}
\bar{Q}_{(n)}={Q_{+(n)}^{(w)}+Q_{-(n)}^{(w)} \over 2} \ , \qquad \Delta Q_{(n)}=Q_{+(n)}^{(w)}-Q_{-(n)}^{(w)} ~.
\end{eqnarray}

\begin{figure}[htbp]
\begin{center}
 \includegraphics[width=.4\linewidth]{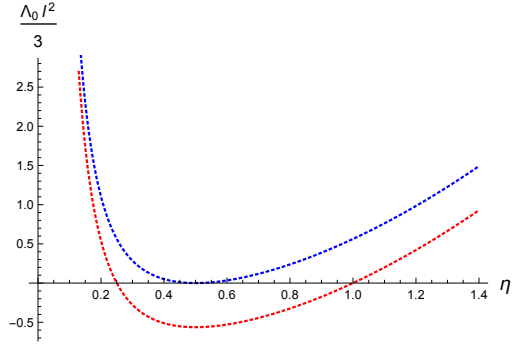}
\vspace{-.0cm}
\caption{\sl The cosmological constant on the bubble as a function of the tension $\eta$. The blue and red curves correspond to $\delta=0$ and $6/10$. }
\label{FigLambda}
\end{center}
\end{figure}

Now, let us discuss the cosmological constant for the four dimensional bubble universe. By comparing with the Friedmann equation, we can read off it from the second term on the right hand side of \eqref{GeneralEQ} in dimensionful coordinates (note that this equation is expressed by Euclidean time),
\begin{eqnarray}
{\alpha^2 \over \gamma^2 } l^2 =   \eta^2-{1+\delta^2 \over 2(1-\delta^2)}+{1\over 16\eta^2}  ={\Lambda^{(4)}l^2\over 3 }~,  \label{CCdef}
\end{eqnarray}
where we defined $\delta={l_-/ l_+}$. The minus sign in the second terms of \eqref{CCdef}, originating from background of the anti-de Sitter space, is important. If we assume de Sitter spaces in five dimensions for both sides of the bubble, the sign becomes plus which make it impossible to take the cosmological constant to be zero. From figure \ref{FigLambda}, we see that the cosmological constant on the bubble depends on the tension $\eta$ and in a wide range of parameter spaces it is positive. 
However, note that as we will see below, not all the tension can be realized under the bubble nucleation: For the fixed value of catalyst, there is the minimum allowed value of $\eta$, below which there is not a bounce solution for the decay (one can check that by numerical computations for explicit models). When $\Lambda^{(4)}<0$, one can also explicitly check that there is no bounce solution for the equation \eqref{GeneralEQ} in the most of $\eta$ (there might be a very small allowed window around the critical value of $\eta$). Hence, the four dimensional AdS space cannot be created by this catalysis. So the minimum cosmological constant obtained by balancing the tension of the bubble and background AdS radius is $\Lambda^{(4)}=0$. In this case, the tension has to satisfy the following relation,
\begin{eqnarray}
\bar{\sigma}^2_{\rm cr}-{1\over 2} \Big( {1\over l_+^2}+{1\over l_-^2}\Big)+{1\over 16\bar{\sigma}^2_{\rm cr}} \Big( {1\over l_+^2}-{1\over l_-^2}\Big)^2=0~.
\end{eqnarray}
We call the bubble satisfying this condition as the critical bubble and denote its tension $\bar{\sigma}_{\rm cr}$. By solving this equation, we obtain the critical value of the tension as
\begin{eqnarray}
\bar{\sigma}_{\rm cr}={1\over 2} \Big({1\over l_-}-{1\over l_+} \Big)\ , \qquad \eta_{\rm cr}={1\over 2}\sqrt{l_+-l_-\over l_++l_-}~. \label{CriticalTension}
\end{eqnarray}
Adopting the same notation in \cite{Dan}, let us introduce a parameter $\epsilon$ representing deviation from the critical value, 
\begin{eqnarray}
\bar{\sigma}=\bar{\sigma}_{\rm cr}(1-{\epsilon})\, , 
\end{eqnarray}
and approximate the right hand side of \eqref{CCdef} a few order in $\epsilon$
\begin{eqnarray}
\Lambda^{(4)}\simeq {6\epsilon \over l_+l_-}+3\epsilon^2 \Big({1\over l_-^2}+{1\over l_+^2}+{1\over l_-l_+} \Big)~.
\end{eqnarray}
Interestingly, by taking $l_+\to \infty$ (Minkowski limit), the leading order of the cosmological constant becomes zero. Hence, to make fine-tuning a bit mild, we assume a metastable Minkowski vacuum for the original geometry and study its decay process to AdS vacuum. It is worthy noting that if we naively take Minkowski vacua for both side of the bubble, which corresponds to $l_\pm \to \infty$, the parameters become $\alpha=1$ and $\gamma=\bar{\sigma}$, thus the cosmological constant given by the scale of the tension, $\Lambda^{(4)}=3\bar{\sigma}^2$. In this case, $G_4$ and $\Lambda^{(4)}$ are roughly expressed by the same scale, namely the Planck scale, which cannot be acceptable in phenomenological point of view. In this scene, we assume AdS (or Minkowski for outside) spacetimes for both sides and study bubbles with tension near the critical value. Also, to make the four dimensional Newton constant finite, we define $G_4={G_5/ l}$.

The quintessence in five dimensions can be interpreted as the one in four dimensions on the bubble. For example, a contribution to the Friedmann equation can be given by ${1/ \widetilde{R}^{4w+4}}$ in five dimensional terminology. This should correspond to ${1/ \widetilde{R}^{3w_{(4)}+3}}$ in four dimensions. Thus, we get
\begin{eqnarray}
w_{(4)}\, \iff \ {1\over 3}(4w+1)~.
\end{eqnarray}
We find that quintessence parameter $w$ in five dimensions corresponds to the states on the bubble as follows\footnote{In four dimensions, $w_{(4)}=-1,0,1/3$ corresponds to the cosmological constant, matter and radiation respectively.}. 
\begin{equation}
 w= \left \{ 
 \begin{array}{l} 
 -1 \quad  \mbox{cosmological constant } \\ 
  -{1\over 4} \quad  \mbox{matter  \qquad \qquad \qquad \qquad (on the bubble)}\\ 
   0 \qquad \mbox{radiation  } 
 \end{array} 
 \right. ~.
\end{equation} 
In this ways, parameter range $-1\le w \le 0$ covers all the states in four dimensions. Thus, below we consider only this parameter range for $w$ in five dimensions.

\subsection{Calculation of the bounce action}

In this subsection, we quickly review how to compute the bounce action. Recently, Gregory, Moss and Withers showed how to treat singularities on bounce solutions \cite{Gregory14}. Here, we outline the formulae without showing the details. To extract the contributions from the singularities, let us decompose the spacetime into two parts. Suppose that there are several singularities on the solution labeled by $i$. We denote the neighborhood of the singularities ${\cal B}=\sum_i {\cal B}_i$ and whole spacetime ${\cal M}$. Subtracting the singular parts, we obtain ${\cal M}-{\cal B}$. In the same way, the action can be decomposed into two parts, $I= I_{ {\cal M}-{\cal B} } +I_{\cal B}$. Let us consider the first contribution
\begin{eqnarray}
I_{ { \cal M}-{ \cal B} } = -{1\over 16\pi G_5} \int_{ {\cal M}-{  \cal B} } R -\int_{ {\cal M}-{\cal B}}{\cal L}_m+{1\over 8\pi G_5}  \int_{\partial ({\cal M}-{\cal B})}K \ . \label{MminusB}
\end{eqnarray}
This can be further divided into three parts, namely the contributions from in and outside of the bubble (we denote ${\cal W}$) and bubble itself, 
\begin{eqnarray}
  I_{ {\cal M}-{\cal B}}=I_-  +I_+  +I_{\mathcal W}\, .
\end{eqnarray}
The action on the wall can be given by 
\begin{eqnarray}
  I_{ \mathcal W}=-\int_{\mathcal W}     {\mathcal L}_m=\int_{\mathcal W} \sigma \, .
\end{eqnarray}
Also, the curvature tensor of five dimensions can be decomposed in terms of the four dimensional curvature as follows: 
\begin{equation}
R={}^{(5)}R-K^2+K_{ij}^2-2\nabla_i (u^i\nabla_j u^j)+2\nabla_j(u^i\nabla_i u^j) \, ,
\end{equation}
where $u^{j}$ is the derivative of coordinates with respect to the proper time on the bubble. By exploiting these expressions,  the action can be written as
\begin{eqnarray}
  I_\pm=-\frac{1}{8\pi G_5}\int_{\mathcal W}K_\pm+\frac{1}{8\pi G_5}\int_{\mathcal W} n_{\pm j}u^i\nabla_i u^j\, ,
\end{eqnarray}
where $n^\mu$ is the normal vector perpendicular to the bubble and satisfies the condition, $1=g_{\mu \nu}n^\mu n^\nu$. Hence, the action \eqref{MminusB} can be expressed as 
\begin{eqnarray}
  I_{ {\cal M}-{\cal B}}&=&\int_{\mathcal W}\sigma-{4 \over 3}\int_{\mathcal W} \sigma-{1\over 16\pi G_5} \int_{\mathcal W}(f^\prime_+\dot{\tau}_+ -f^\prime_-\dot{\tau}_-)\,  \\
  &=&-{1\over 3}\int_{\mathcal W}\sigma-{1\over 16\pi G_5}\int_{\mathcal W}(f^\prime_+\dot{\tau}_+-f^\prime_-\dot{\tau}_-)\, .
\end{eqnarray}

Next, we show contributions to the bounce action from the singularities on the solution. According to \cite{Gregory14,Gregory15}, these are given by the entropy of the horizons. (See, for example, appendix of \cite{KO2}.)
\begin{eqnarray}
I_{ {\cal B} }&=& -{1\over 16\pi G_5} \int_{\cal B} R -\int_{  {\cal B}}{\cal L}_m +{1\over 8\pi G_5} \int_{ \partial {\cal B}} K \,  \\
&=&-{1\over 4G_5}\sum_i {\cal A}_i\, .
\end{eqnarray}
The bounce action $B$ for the decay process is given by subtraction of the action $I_0$ for the original configuration,
 \begin{equation}
 B=I_{B}-I_{0} \, .
 \end{equation}

Finally, let us comment on a contribution to the bounce action from the singularity at the origin of the quintessence solution. For the solution with $-1/2< w \le 0$ and $r_{BH}=0$, there is an event horizon surrounding the origin which allows us to consider only outside region of it to calculate the bounce action. However, when we study the solution with $-1< w \le -1/2$ and $r_{BH}=0$, there is a naked singularity which can contribute to the bounce action. However, by using the method shown in \cite{Gregory14}, one can show that this singularity does not contribute: To show that, let us turn on $r_{BH}$ and introduce a small black hole horizon, which can be regarded as a regulator for the calculation. In this case, by the method in \cite{Gregory14}, we find that its contribution is proportional to the area of the horizon. In the limit  $r_{BH}\to 0$, this vanishes. Therefore, we do not have to take into account the singularity at the origin.

\section{Catalysis induced by quintessence}

In this section, we illustrate catalytic effects by discussing the freezing and thawing models separately. We investigate behavior of the bounce action as a functions of tension and quantities of the catalysts. We will find that in both cases, the bounce actions becomes smaller compared to that of Coleman-de Luccia, hence the lifetime of a metastable state is much shorter. We will see that there is a tendency that the catalytic effect becomes more important when $w$ approaches to $-1/2$.

\subsection{Catalysis in the freezing model}

\begin{figure}[t]
\begin{center}
  \includegraphics[width=.45\linewidth]{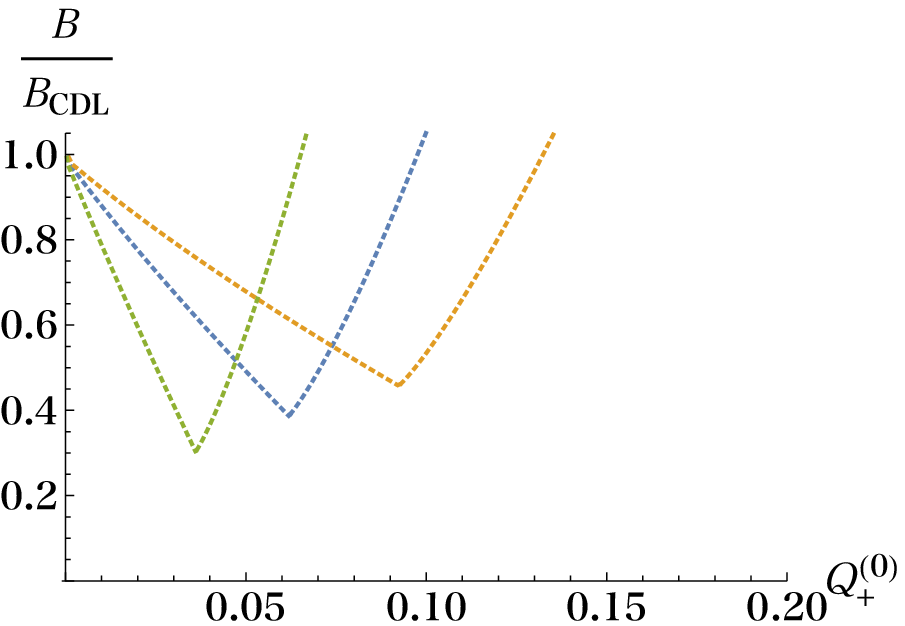}\hspace{1cm} 
 \includegraphics[width=.45\linewidth]{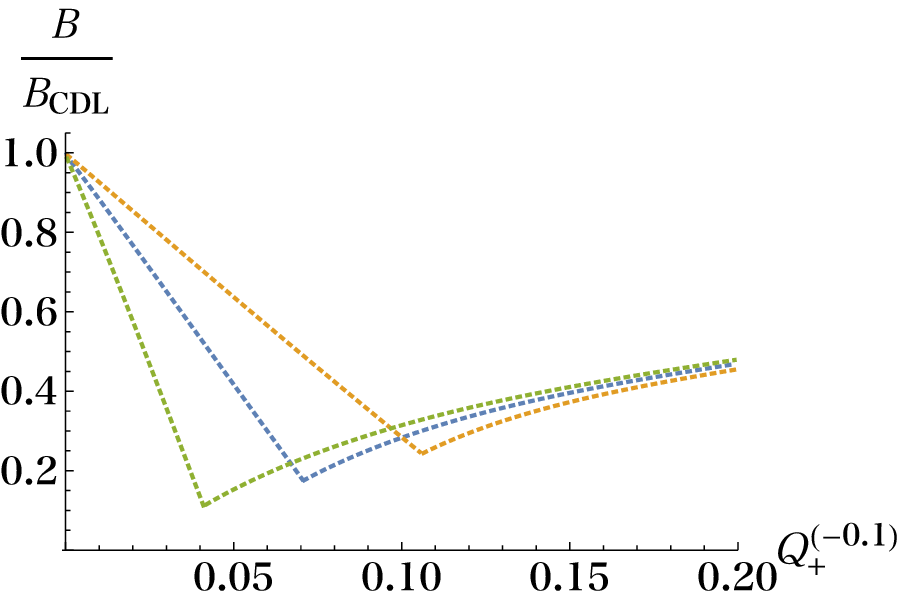} 
 \includegraphics[width=.45\linewidth]{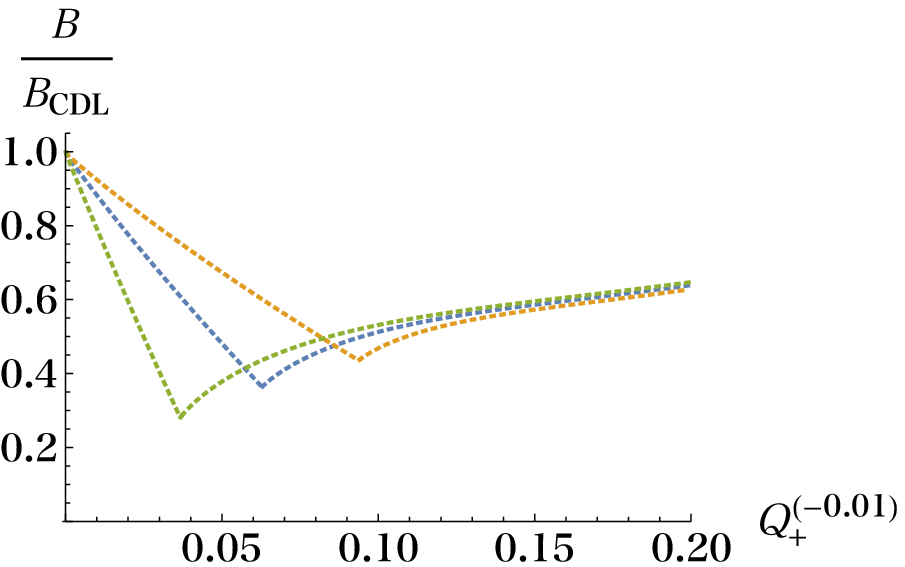} \hspace{1cm}
  \includegraphics[width=.45\linewidth]{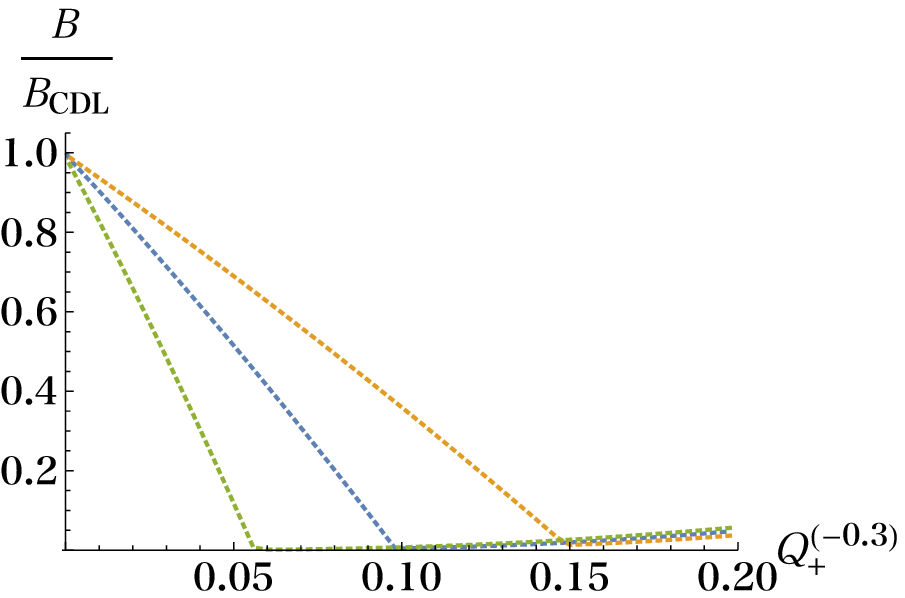} 
\vspace{-.0cm}
\caption{\sl The bounce actions for the quintessence with $w=0$, $-1/100$, $-1/10$ and $-3/10$. The green, blue and orange curves correspond to $\eta =0.15$, $0.2$ and $0.25$. Beyond the minimum values of $Q_+^{(w)}$ for each curve, there is a remnant in the bubble. We refer to this bubble as the saturated bubble. }
\label{FigFreezing}
\end{center}
\end{figure}

As the first example of the catalysis, let us consider the freezing model: In the early stage of the universe, $w$ is close to zero and gradually shits to smaller value (see figure \ref{FigOurAssumption}). Thus, we consider catalysis induced by quintessence with $w\sim 0$.  We assume the decay of metastable Minkowski vacuum to AdS one. So, $f_\pm$ in this case are given by
\begin{eqnarray}
f_+=1-{Q^{(w)}_+\over \tilde{r}^{4w+2}} \ , \qquad f_-=1-{Q^{(w)}_-\over \tilde{r}^{4w+2}}+\Big({\gamma \over l_- \alpha} \Big)^2 \tilde{r}^2~.
\end{eqnarray}
As mentioned in general arguments, the equation of motion for the bubble radius $\widetilde{R}$ (in Euclidean time) is given by 
\begin{eqnarray}
(\dot{\widetilde{R}})^2=1-\widetilde{R}^2 - \Big(\bar{Q}+{\Delta Q\over 8\eta^2} \Big){1  \over \widetilde{R}^{4w+2}}-\Big({(1+4\eta^2)\alpha \Delta Q \over 16\eta^2} \Big)^2 {1 \over \widetilde{R}^{8w+6}}~. \label{EOM3}
\end{eqnarray}
As an illustration, we show the bounce action for the cases, $w=0$, $-1/100$, $-1/10$ and $-3/10$ in the figure \ref{FigFreezing}. Remarkably, the catalytic effect is enhanced for small value of $w$. Therefore, in an explicit freezing model, the catalysis becomes more important as the time goes by. From the figure, we find that for small value of $Q_+$, the bounce action is monotonically decreasing function and eventually reaches the minimum. We call the bubble at this point the saturated bubble. Above this minimum, the property of the bubble is changed and in fact, there exists a remnant inside the bubble.

What happen if we take smaller value of $w$? The catalytic effect can become strong, however the inner horizon approaches to the size of the bubble since  it is given by $\tilde{r}= Q_+^{1/(4w+2)}$. Clearly at $w=-1/2$, it diverges, which means that the horizon goes to infinity. This should be regard as a cosmological horizon. So with $w<-1/2$, the model is similar to the thawing type, we will discuss it in the next subsection.

\subsection{Catalysis in the thawing model}

\begin{figure}[t]
\begin{center}
 \includegraphics[width=.45\linewidth]{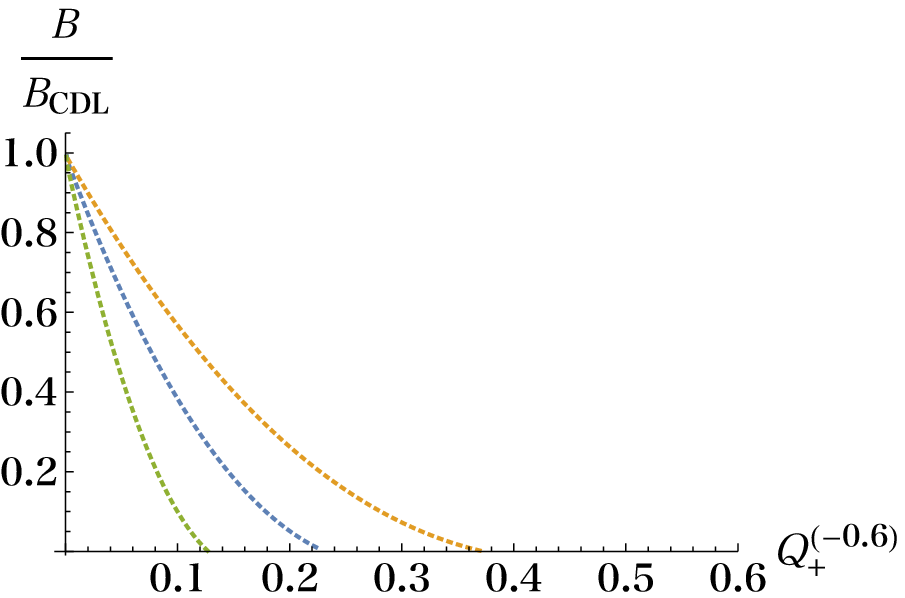} \hspace{1cm}
 \includegraphics[width=.45\linewidth]{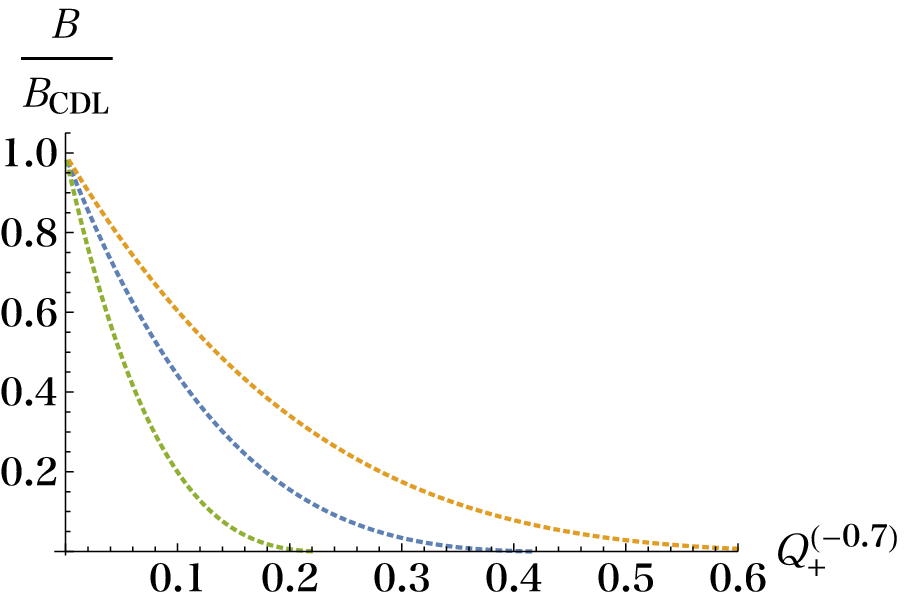}
  \includegraphics[width=.45\linewidth]{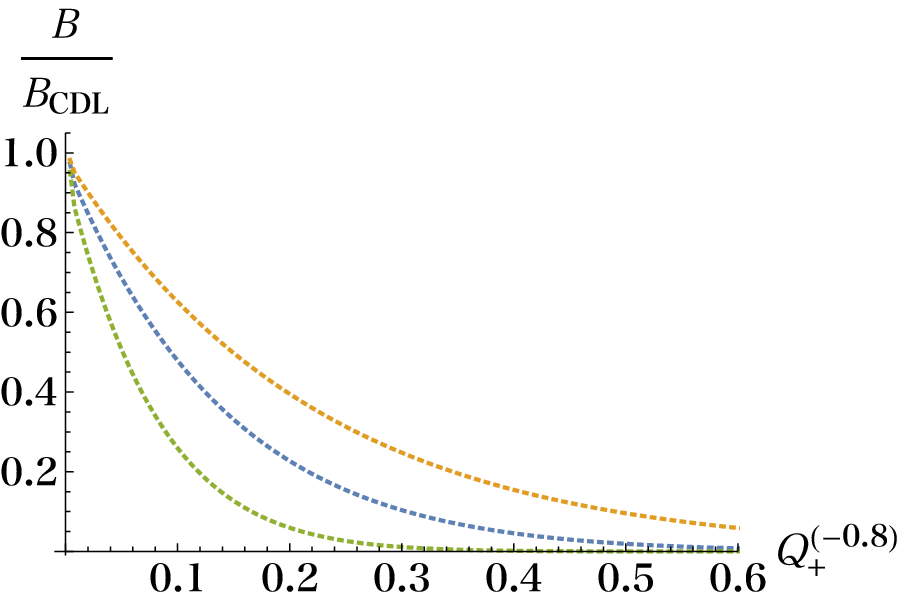} \hspace{1cm}
   \includegraphics[width=.45\linewidth]{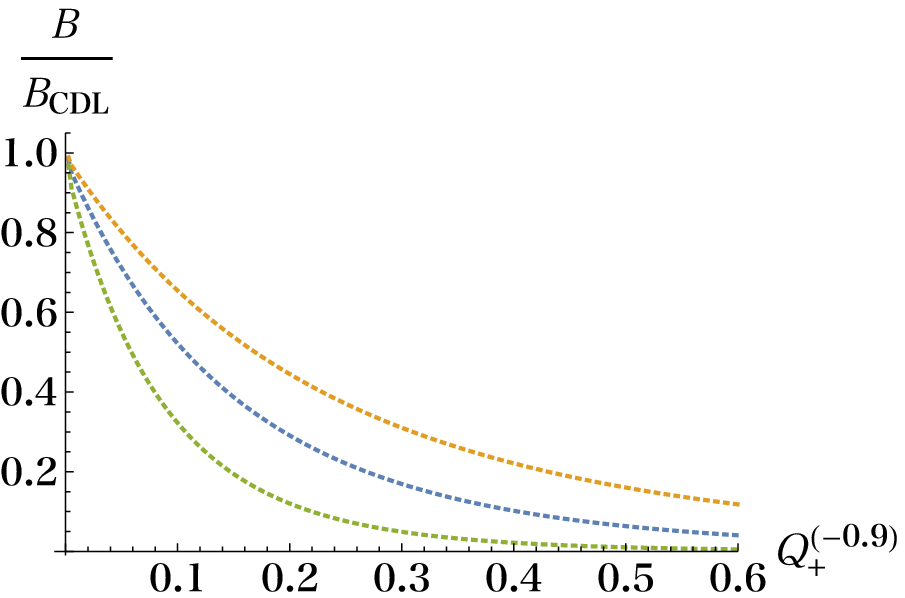}
\vspace{-.1cm}
\caption{\sl The bounce actions for the thawing model with $w=-6/10$, $=-7/10$, $=-8/10$ and $=-9/10$. The green, blue and orange curves correspond to $\eta=0.15$, $\eta=0.2$ and $\eta=0.25$. }
\label{FigThawing}
\end{center}
\end{figure}

As the second example of the catalysis caused by quintessence, let us study the thawing type behavior of $w$. In this model, at the early stage of the universe $w$ is relatively close to $-1$. In this subsection, we naively assume $-1< w<-1/2$. When $w\simeq -1$, the geometry becomes similar to the de Sitter spacetime because  $f_+$ behaves like
\begin{equation}
f_+=1-Q_+ \widetilde{R}^a \ , \qquad a \simeq 2 ~.
\end{equation}
Again, we assume the metastable Minkowski vacuum, namely $\Lambda^{(5)}_+$ is zero,  as an initial state. Since this $Q_+$ plays a similar role to the cosmological constant of the de-Sitter space, we find that smaller values of $Q_+$ are energetically favorable. Thus, we expect that catalytic effect works in this model as well. Similarly, the geometry inside the bubble can be given by  
\begin{eqnarray}
f_-=1-Q_- \widetilde{R}^a +\Big({\gamma \over l_- \alpha} \Big)^2 \widetilde{R}^2 ~.
\end{eqnarray}
As mentioned above, smaller $Q_-$ is energetically favorable. The equation of motion for the bubble is given by \eqref{EOM3}. As an illustration, we show the bounce action for the model with $w=-6/10$, $=-7/10$, $=-8/10$ and $=-9/10$ in the figure \ref{FigThawing}. Again, one find that the bounce action becomes smaller as the parameter $w$ approaches to $-1/2$. 

\section{Realization of Inflation and dark energy on the bubble}

In the previous section, we studied the catalysis induced by quintessence. Both freezing and thawing models had an enhanced decay rate of metastable vacuum and quintessence played a role of catalyst. In this section, exploiting this understanding, we propose a model realizing inflation and the dark energy on the expanding bubble universe. On this bubble four dimensional gravity is localized \cite{Dan} due to AdS spaces inside and outside of the bubble. The cosmological constant for the four dimensional theory can be determined by the tension of the bubble and vacuum energies of AdS spaces. Since the tension of bubble is fixed by dynamics of decay process, we can find most probable cosmological constant realized on this decay process. We will search in a wide range of parameter spaces of the theory. Then, we impose the trans-Planckian censorship conjecture (TCC) \cite{TCC} and discuss allowed cosmological constant, which gives us a constraint for the allowed parameter spaces. Since the created bubble itself is stable, as long as the lower energy vacuum is the absolute minimum, all the bubble with positive cosmological constant cannot satisfy TCC. This constraint give us remarkable scenario for realizing four dimensional theory in string theories.

\subsection{Catalytic selection of bubble universe}

\begin{figure}[htbp]
\begin{center}
\includegraphics[width=.5\linewidth]{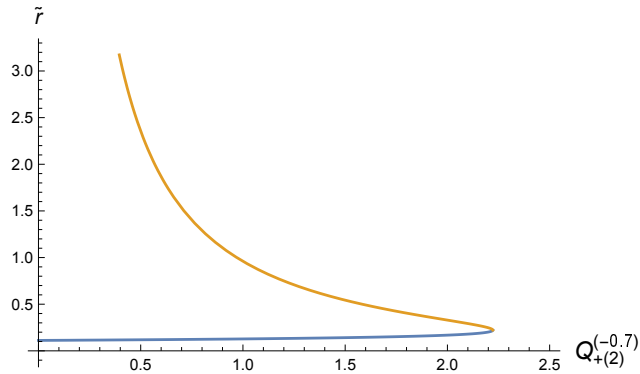}
\vspace{-.1cm}
\caption{\sl Outer (the upper curve) and inner (the lower curve) horizons of the geometry before the transition. We choose $w_1=-1/10$, $w_2=-7/10$ and $Q^{(-0.1)}_{+(1)}=3/100$. Around $Q_{+2}^{(-0.7)}\sim 2.2$ they coincide with each other.}
\label{FigHorizons}
\end{center}
\end{figure}

Let us include freezing and thawing types of quintessence simultaneously which eventually play roles of the dark energy and inflaton. Then, we assume that  they have $w_1\simeq -1$ and $w_2\simeq 0$ when a bubble is created. Note that the most economical earlier example is the one discussed in \cite{QuintessentialInflation,Hashiba} where single scale field plays not only the inflaton for the primordial inflation but also the quintessential dark energy in the late stage. However, this model cannot satisfy the conditions coming from the distance conjecture \cite{Swampland} and the trans-Planckian censorship conjecture \cite{TCC}, so we introduce two kinds of quintessence fields and consider a model circumventing the swampland conjectures.  In the same way as previous section, we study the decay of Minkowski vacuum to AdS vacuum for the sake of simplicity. The inside and outside geometries that we will study are as follows: 
\begin{eqnarray}
f_+= 1-{Q_{+(1)}^{(w_1)}\over \tilde{r}^{4w_1+2}}-{Q_{+(2)}^{(w_2)}\over \tilde{r}^{4w_2+2}} \ ,\qquad f_-= 1-{Q_{-(1)}^{(w_1)}\over \tilde{r}^{4w_1+2}}-{Q_{-(2)}^{(w_2)}\over \tilde{r}^{4w_2+2}}+\Big( {\gamma \over l_-\alpha} \Big)^2\tilde{r}^2~.
\end{eqnarray}
The equation of motion for the bubble connecting two geometries at $\tilde{r}=\widetilde{R}$ is given by
\begin{eqnarray}
(\dot{\widetilde{R}})^2&=&1-\widetilde{R}^2 - \Big(\bar{Q}_{(1)}+{\Delta Q_{(1)}\over 8\eta^2} \Big){1  \over \widetilde{R}^{4w_1+2}}- \Big(\bar{Q}_{(2)}+{\Delta Q_{(2)}\over 8\eta^2} \Big){1  \over \widetilde{R}^{4w_2+2}}\nonu \\
&&-{1\over 16\eta^2} \Big( {l\alpha \over \gamma} \Big)^2 \Big( {\Delta Q_{(1)}\over \widetilde{R}^{4w_1+3} } + {\Delta Q_{(2)}\over \widetilde{R}^{4w_2+3} } \Big)^2~.
\end{eqnarray}
The original geometry before bubble nucleation has two horizons. In the figure \ref{FigHorizons}, we show an example of inner and outer horizons for the parameter choice,  $w_1=-1/10$, $w_2=-7/10$ and $Q^{(-0.1)}_{+(1)}=3/100$. For fixed $Q_{+(1)}^{(w_1)}$ and $w_{1,2}$, there is the maximum value of $Q_{+2}^{(w_2)}$ where two horizons coincide. In the figure, it exists around $Q_{+2}^{(-0.7)}\sim 2.2$.

\begin{figure}[t]
\begin{center}
\includegraphics[width=.5\linewidth]{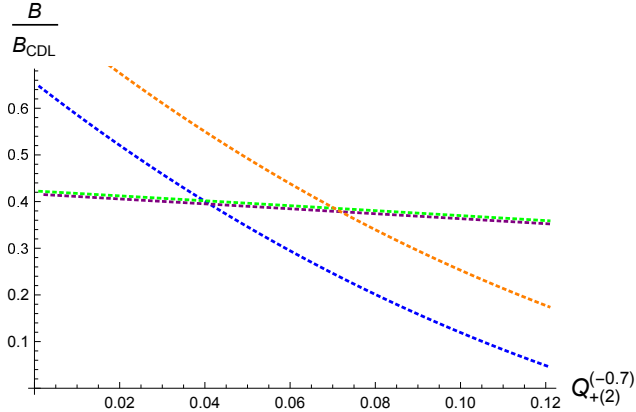}
\vspace{-.1cm}
\caption{\sl The numerical calculation of the bounce action for the parameter choice $\eta=0.2$, $Q_{+(1)}^{(-0.1)}=3/100$, $w_1=-1/10$ and $w_2=-7/10$. The blue curve corresponds to the bubble without remnant, namely $Q_{-(1)}^{(-0.1)}=0$ and $Q_{-(2)}^{(-0.7)}=0$. The orange and green curves have two types remnants, $Q_{-(1)}^{(-0.1)}=1/100$, $Q_{-(2)}^{(-0.7)}=1/100$, $Q_{-(1)}^{(-0.1)}=3/100, Q_{-(2)}^{(-0.7)}=1/100$, respectively. The purple curve includes only one remnant  $Q_{-(1)}^{(-0.1)}=3/100, Q_{-(2)}^{(-0.7)}=0$. 
}
\label{FigBounce1}
\end{center}
\end{figure}

Now, we are ready to calculate the bounce action for the decay of the Minkowski vacuum to AdS vacuum catalyzed by two quintessence fields. Especially, we focus on the following parameter choice, $\eta=0.2$, $Q_{+(1)}^{(-0.1)}=3/100$, $w_1=-1/10$ and $w_2=-7/10$. We show the ratio of the bounce action to that of CDL for various choices of  remnants $Q_{- (1,2)}^{(w_{1,2})}$ in the figure \ref{FigBounce1}. For small value of $Q_{+(2)}^{(-0.7)}$, the dominant decay process is given by $Q_{-(1)}^{(-0.1)}=3/100$ and $Q^{(-0.7)}_{-(2)}=0$ (the purple curve), which means that the bubble with remnant of quintessence of $w_1$ is the most probably possibility. On the other hand, when $Q_{+(2)}^{(-0.7)}$ is larger than about $0.04$, the bubble without remnants, namely $Q_{-(1)}^{(-0.1)}=0$ and $Q_{-(2)}^{(-0.7)}=0$, dominate the process (the blue curve). The former region, the catalytic effect induced by quintessence of $w_2$ is dominant while the latter is that of $w_1$. Since $w_2$ is smaller than ${-1/2}$, it behaves like positive cosmological constant which clearly enhances the decay rate of the vacuum in any region. On the other hand, as for $Q_{+(1)}^{(w_1)}$, behavior is slightly different. Turning on $Q_{+(1)}^{(-0.1)}$ makes event horizon in the geometry which create a singularity of the bounce solution. This enlarges the bounce action and forbids decay of $w_1$ quintessence. However, by taking $Q_{+(2)}$ large, the catalytic effect overcomes that and allows us to decay both of quintessence in region of large value of $Q_{+(2)}$. 

\begin{figure}[t]
\begin{center}
\includegraphics[width=.45\linewidth]{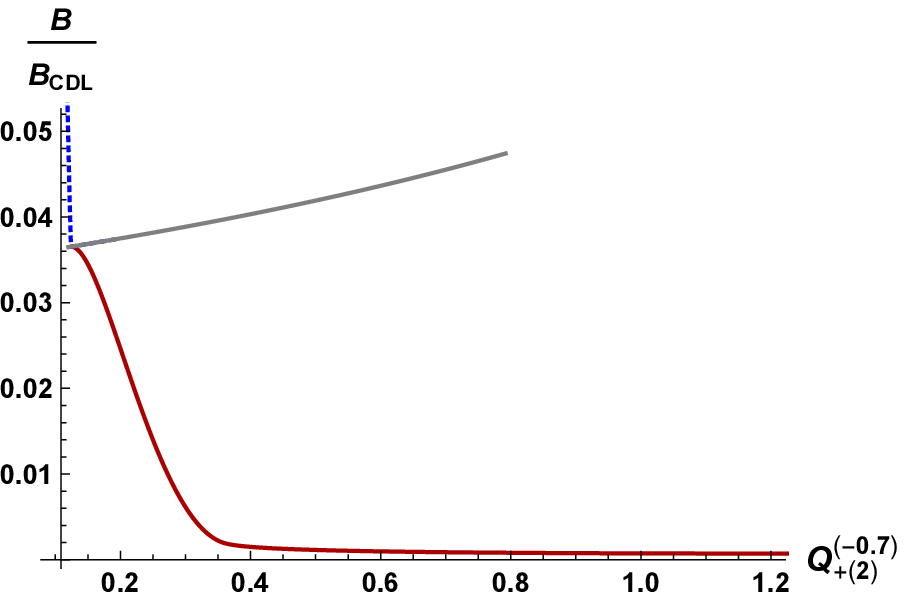}\hspace{1cm}
\includegraphics[width=.45\linewidth]{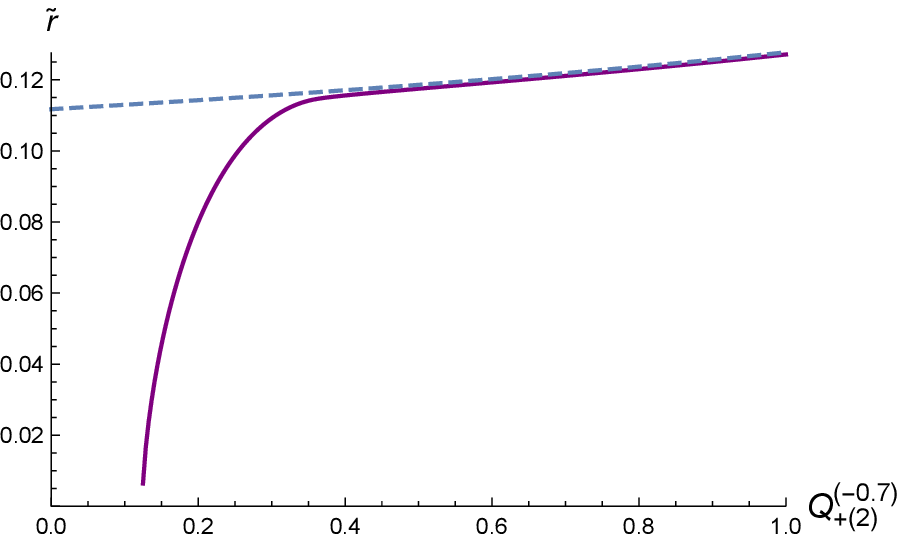}
\vspace{-.1cm}
\caption{\sl Left panel: The bounce actions for the saturated bubble in which there is a remnant. We took $\eta=2/10$, $Q^{(-0.1)}_{+(1)}=3/100$, $w_1=-1/10$ and $w_2=-7/10$.
The gray curve corresponds to the bubble with $Q^{(-0.1)}_{-(1)}=0$ and nonzero $Q^{(-0.7)}_{-(2)}$ while the brown curve corresponds to that with $Q^{(-0.7)}_{-(2)}=0$ and nonzero $Q^{(-0.1)}_{-(1)}$. The blue curve are connected to the end point of the one in figure 6. Right panel: purple curve is the position of inner horizon inside the bubble. The doted blue line is the that of original geometry before the transition. }
\label{FigRemnants}
\end{center}
\end{figure}

Beyond $Q^{(-0.7)}_{+(2)}\sim 0.12$, the dominant contribution is given by the saturated bubble discussed in the previous section. In other words, the remnant has to be required to satisfy the equation of motion. The remnant should be $Q_{-(1)}^{w_1}$ or $Q_{-(2)}$. To see which remnant yields the dominant contribution, in the figure \ref{FigRemnants}, we show two bounce actions corresponding to two choices of  remnants for the parameter choice, $\eta=2/10$, $w_1=-1/10$, $w_2=-7/10$ and $Q^{(-0.1)}_{+(1)}=3/100$. For the gray curve, we assumed  $Q^{(-0.1)}_{-(1)}=0$ and nonzero $Q^{(-0.7)}_{-(2)}$. The nonzero value is determined by solving conditions for the saturated bubble. On the other hand, the blue curve is opposite choice with $Q^{(-0.7)}_{-(2)}=0$ and nonzero $Q^{(-0.1)}_{-(1)}$. Clearly, the bounce action for the blue curve is smaller than the other.

Let us comment on behavior of the blue curve around $Q_{+(2)}^{(-0.7)}\sim 0.3$. This drastic change of the behavior can be understood from the contribution of the inner horizon after the transition. As one can see in the right panel of the figure \ref{FigRemnants}, around $Q_{+(2)}^{(-0.7)}\sim 0.3$, the inner horizon after the transition approaches to the size of the horizon before the transition. Thus, the contributions from the singular part of the bounce solution before and after the transition becomes almost the same, which reduces the bounce action.

\begin{figure}[t]
\begin{center}
\includegraphics[width=.45\linewidth]{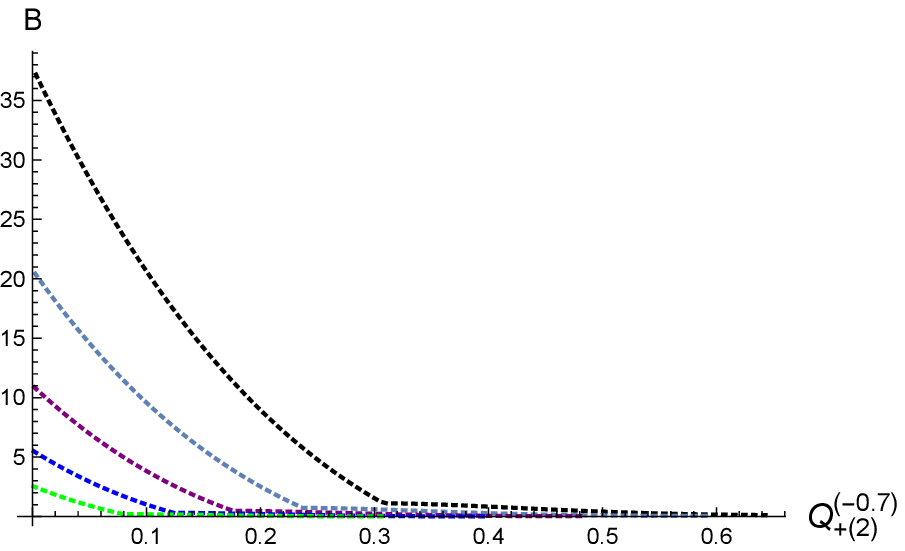}\hspace{1cm}
\includegraphics[width=.45\linewidth]{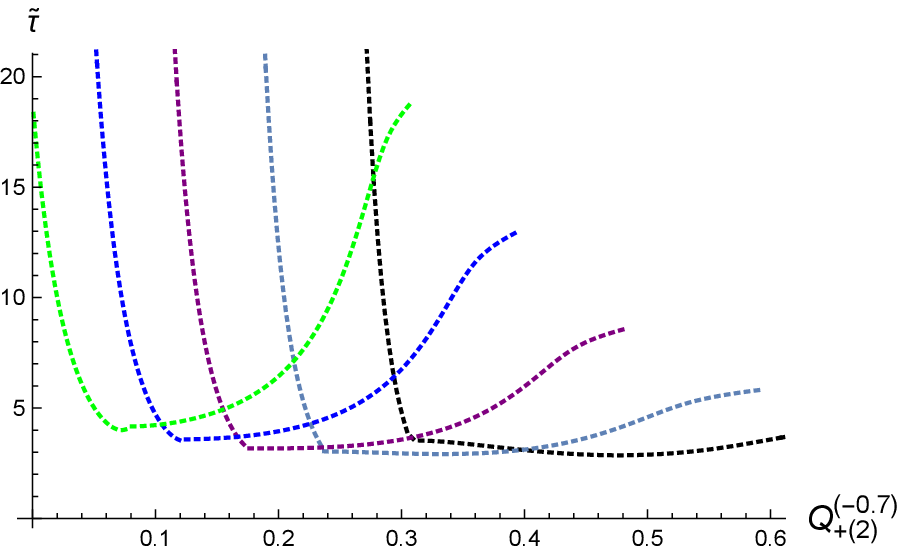}
\includegraphics[width=.45\linewidth]{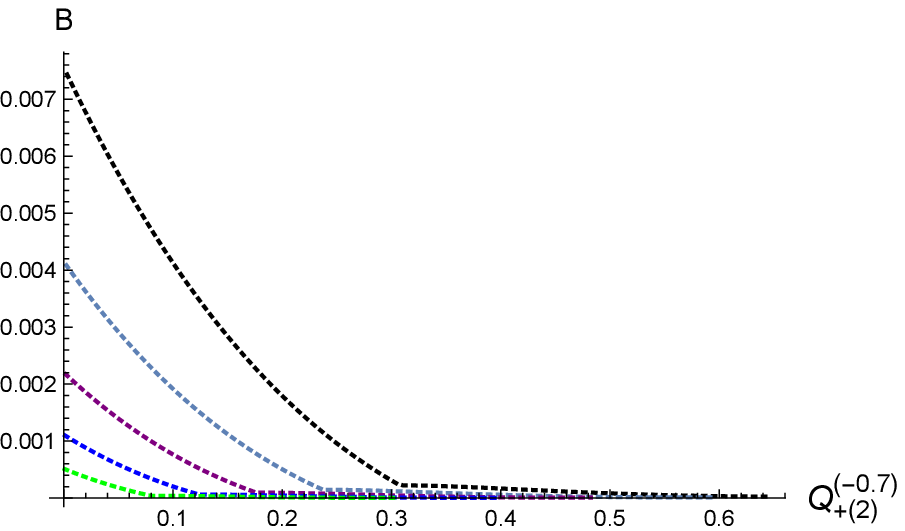}\hspace{1cm}
\includegraphics[width=.45\linewidth]{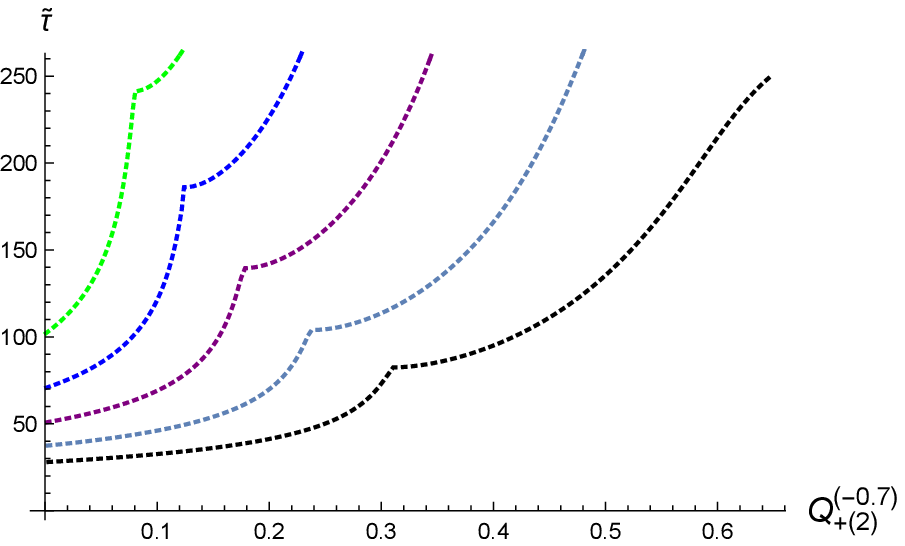}
\vspace{-.1cm}
\caption{\sl We choose the parameters $Q_{+(1)}^{(-0.1)}=3/100$, $w_1=-1/10$, $w_2=-7/10$ and the green, blue, purple, light-blue and black curves correspond to $\eta=0.18$, $0.2$, $0.22$, $0.24$ and $0.26$. In the upper two panels, we set $l^3/G_5=50$. In this case, the pre-factor significantly contribute and bubbles with larger $B$ becomes the most probable decay. On the other hand, in the lower two panels, we set $l^3/G_5=1/100$. In this case, the values of the bounce action becomes larger compared to the previous case, the dominant contribution is given by the bubble with smaller bounce action. }
\label{FigHybrid1}
\end{center}
\end{figure}

\begin{figure}[t]
\begin{center}
\includegraphics[width=.45\linewidth]{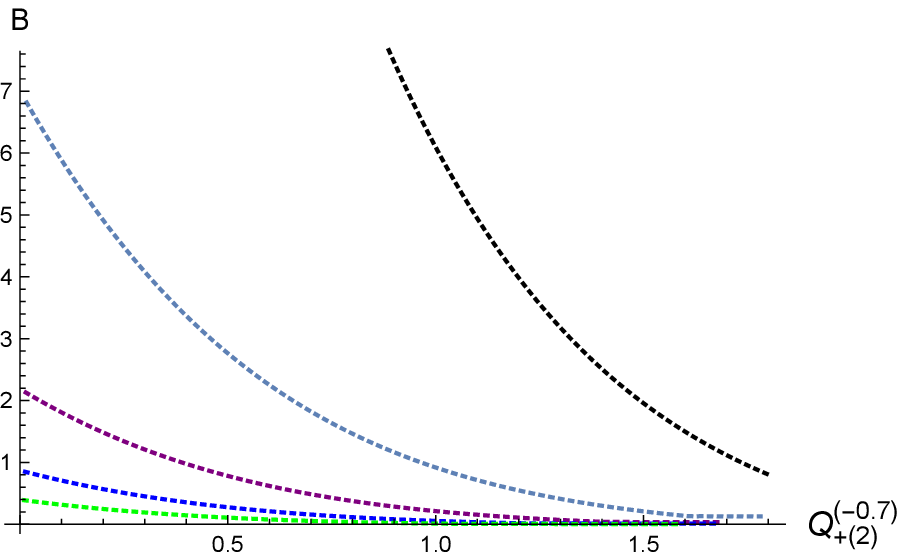}\hspace{1cm}
\includegraphics[width=.45\linewidth]{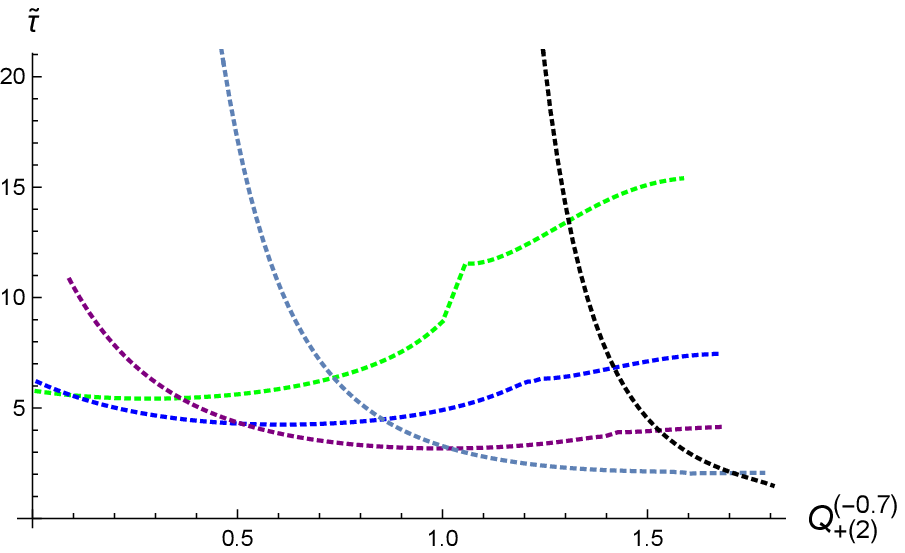}
\vspace{-.1cm}
\caption{\sl The bounce actions for larger values of tension. The green, blue, purple, light-blue and black curves correspond to $\eta=0.39$, $041$, $0.43$, $0.45$ and $0.47$. We took $l^3/G_5=1/100$, $Q^{(-0.1)}_{+(1)}=3/100$, $w_1=-1/10$ and $w_2=-7/10$.}
\label{FigHybrid2}
\end{center}
\end{figure}

\begin{figure}[htbp]
\begin{center}
 \includegraphics[width=.45\linewidth]{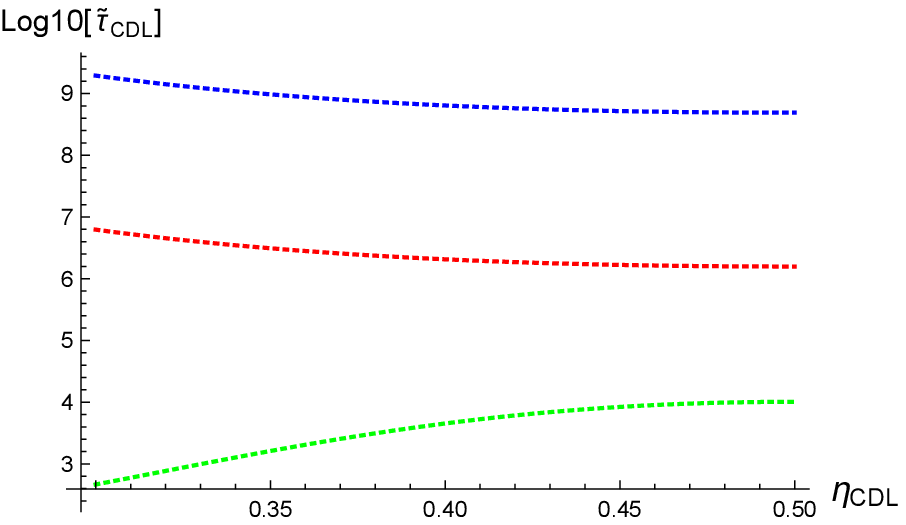}
\vspace{-.0cm}
\caption{\sl Numerical estimation of lifetime for the CDL solution.  Blue, red and green curves correspond to $l^3/G_5=1/100$, $1/10$ and $50$. }
\label{CDLlifetime}
\end{center}
\end{figure}

As we have shown in the figure \ref{FigBounce1} and \ref{FigRemnants}, the bounce actions become smaller as the parameter $Q_{+(2)}^{(w)}$ increases, which naively suggest us that the most probable bubble corresponds to that of the smaller action because the decay rate is proportional to the action, $\Gamma \sim e^{-B}$. However when $B$ is very small, the pre-factor coming from one-loop contribution around the bounce solution can have a large contribution. By taking into account them, the lifetime of the vacuum is given by
\begin{eqnarray}
{\tau}\simeq \Big({ 2\pi \over B} \Big)^{1\over 2} R_{ B } e^{B}~,
\end{eqnarray}
where $R_{B}$ is the size of the bubble. For numerical simulation it is useful to define dimensionless lifetime as $\tilde{\tau}={\alpha}\tau/\gamma$. In the figure \ref{FigHybrid1} and \ref{FigHybrid2} we show the bounce action and the lifetime of metastable vacuum as functions of  $Q_{\pm (2)}^{(w)}$ for the parameter choice, $Q^{(-0.1)}_{+(1)}=3/100$, $w_1=-1/10$, $w_2=-7/10$, ${l^3/G_5}=50$ and $1/100$. From the figures, we see that the lifetime of the vacuum is much shorter than that of Coleman-de Luccia in five dimensions, 
\begin{eqnarray}
\widetilde{\tau}_{CDL}\simeq  \Big({2\pi \over B_{CDL}}\Big)^{5/2} \widetilde{R}_B e^{B_{CDL}}~. \label{LTCDL}
\end{eqnarray}
The numerical estimations of the lifetime of CDL solution are shown in the figure \ref{CDLlifetime}. The most probable universe can depend basically on two parameters, $l^3/G_5$ and $Q_{+(2)}^{(w)}$. In general, when $l^3/G_5$ is small, there is a tendency that the higher tension bubble is selected by the catalysis. By taking $l^3/G_5$ to be very small, the tension of the most probable universe approaches to the critical tension \eqref{CriticalTension}. In this sense, for the critical bubble to have the largest probability by catalysis, we have to finetune the parameter $l^3/G_5$, which is the price we have to pay to get very small cosmological constant on the four dimensional bubble universe in light of the catalytic selection.     

\subsection{Trans-Planckian censorship conjecture and catalytic creation of bubble}

 In the previous subsection, we studied a selection of bubble in terms of the catalysis and find that the cosmological constant on the bubble is determined by parameter choice of two vacua and $G_5$. Here, we discuss the selection from a slightly different viewpoint. In \cite{TCC}, the trans-Planckian censorship conjecture was proposed in the context of the swampland conjectures. Since we are interested in finding a scenario satisfying all conditions coming from a quantum theory of gravity, we would like to impose this condition to the bubble selection.  TCC condition is given by 
\begin{eqnarray}
{R_f \over R_i}< {M_{\rm pl}\over H_f}, 
\end{eqnarray}
where $R$ is the scale factor of the four dimensions and subscripts $i$ and $f$ indicate the initial and final states. This condition yields a strong constraint for the lifetime of the de Sitter vacuum, thus, in our model, this should be imposed on the four dimensional universe on the bubble. To avoid the trans-Planckian problem, the time-period of the inflation, $T$, has to satisfy the following condition
\begin{eqnarray}
T< H^{-1} \log \Big({M_{\rm pl}\over H} \Big)~,
\end{eqnarray}
 where $H$ is the Hubble parameter during the inflation. In our setup, the created bubble exists eternally, as long as the lower energy vacuum is the absolute stable, it eventually violates the TCC condition. This is remarkable: Although two vacua (in and outside) in the five dimensions are AdS spaces and satisfy all swampland conjectures including TCC, some of created bubbles having the positive cosmological constant can violate the TCC condition. We interpret this fact as follows: Since we naively assume that the tension of the bubble $\bar{\sigma}$ is a free parameter in the thin-wall approximation which comes form the shape of the potential. However, in the low energy theory arising from a consistent theory of gravity, allowed value of $\bar{\sigma}$ is limited, which indicates a constraint for the potential shape in AdS space. Hence, the bubble with positive cosmological constant cannot be created under the decay process. Moreover, as mentioned in the previous sections, four dimensional AdS space on the bubble also cannot be created by the decay process of metastable vacuum. Namely, the bounce solution does not exist for this decay. Therefore, combining these two facts, we remarkably conclude that the created bubbles necessarily have vanishing cosmological constant.\footnote{Here, we assume that the lower energy vacuum with $\Lambda^{(5)}_-$ is absolute minimum. If it is a metastable, then there is a chance to circumvent TCC condition by decaying into a lower energy vaccum. }  

\begin{figure}[t]
\begin{center}
 \includegraphics[width=.4\linewidth]{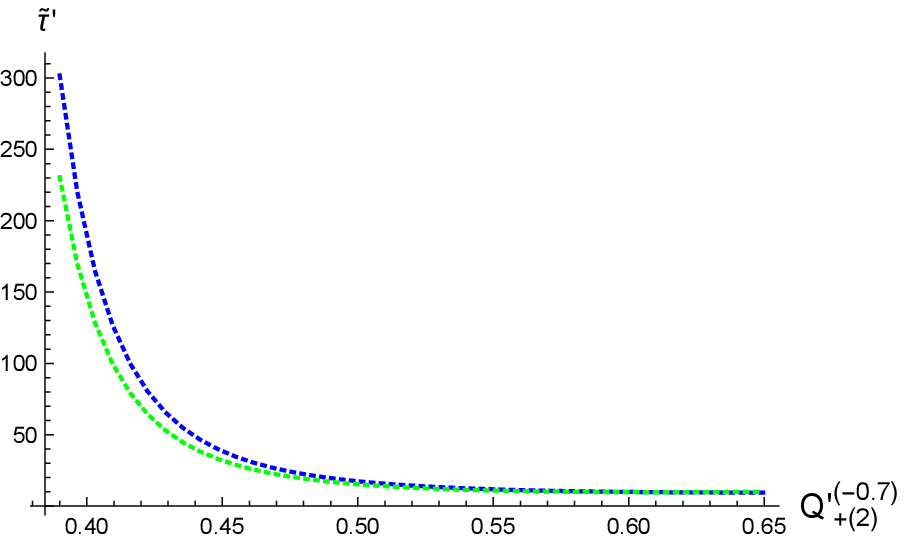}\hspace{2cm}
  \includegraphics[width=.4\linewidth]{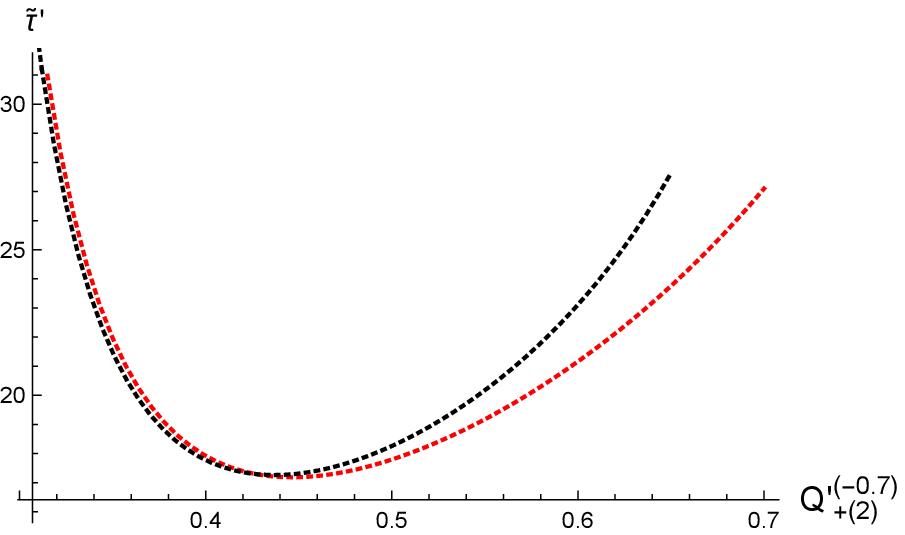}
\vspace{-.0cm}
\caption{\sl Left panel: We choose $l^3/G_5=1/10$, $w_1=-1/10$ and $w_2=-7/10$. The blue and green curves correspond to $Q^{\prime (-0.1)}_{+(1)}=3/100$ and $1/10$. Right panel: The red and black curves are corresponds to $Q^{\prime (-0.1)}_{+(1)}=3/100$ and $1/10$. We took $l^3/G_5=1/100$, $w_1=-1/10$ and $w_2=-7/10$. }
\label{VanishingLambda}
\end{center}
\end{figure}

To see the most probable universe under this assumption, let us consider the equation of motion for the bubble 
\begin{eqnarray}
(\dot{\widetilde{R^\prime}})^2&=&1- \Big(\bar{Q}^\prime_{(1)}+{\Delta Q^\prime_{(1)}\over 8\eta^2} \Big){1  \over \widetilde{R}^{\prime 4w_1+2}}- \Big(\bar{Q}^\prime_{(2)}+{\Delta Q^\prime_{(2)}\over 8\eta^2} \Big){1  \over \widetilde{R}^{\prime 4w_2+2}}\nonu \\
&&-{1\over 16\eta^2} \Big( {l \over \gamma} \Big)^2 \Big( {\Delta Q^\prime_{(1)}\over \widetilde{R}^{\prime 4w_1+3} } + {\Delta Q^\prime_{(2)}\over \widetilde{R}^{\prime 4w_2+3} } \Big)^2~.
\end{eqnarray}
Note that $\widetilde{R}^2$ term does not exist in this case. Since we treat $\alpha=0$, it is convenient to redefine dimensionless parameters as follows:
\begin{equation}
\widetilde{R}^\prime ={ R \over \gamma}\  ,\quad\  \quad \tilde{\lambda}^\prime={\lambda \over \gamma}\  ,\quad \  \quad \tilde{ \tau}^\prime={ \tau  \over \gamma} \  ,\quad\  \quad Q_{ \pm(n)}^{\prime (w)}=\Big({1 \over \gamma} \Big)^{4w+2}q_{\pm(n)}~.
\end{equation}
As in the previous subsection, we can calculate the bounce action and the lifetime of the metastable vacuum. Numerical estimations are shown in figure \ref{VanishingLambda}. Again, the lifetime is much shorter than that of CDL shown in \eqref{LTCDL}. 

\subsection{Application to cosmology}

Here, we will discuss expansion of the bubble after the nucleation by the catalysis. We use the Minkowski times and consider the evolution of the radius ${R}$. We focus on the critical bubble on which the cosmological constant is vanishing. As mentioned, the normalization of \eqref{normalization} is not appropriate since $\alpha=0$ and $\eta=1/2$, so we write the equation of motion in terms of dimensionful parameters
\begin{eqnarray}
{\dot{{R}}^2\over R^2}=-{1\over R^2}+  {q_{+(1)}\over {R}^{4w_1+4}}+  {q_{+(2)}\over {R}^{4w_2+4}}+{l^2 \over 4} \Big(  {\Delta q_{(1)} \over {R}^{4w_1+4}} + {\Delta q_{(2)} \over {R}^{4w_2+4} }\Big)^2~, \quad \mbox{(for critical bubble)} ~.\label{LateTime}
\end{eqnarray}
We use one of the quintessence fields in our model as an inflaton, namely the one corresponding to $w_1$. Right after the nucleation, it is still $w_1\simeq -1$ since the transition time is quite short-period. As long as $w_1$ is very close to $-1$ the inflation occurs and the radius of the bubble grows exponentially,
\begin{eqnarray}
{R}\propto \exp \Big[{\lambda}\sqrt{q_{+(1)}}  \Big]~.
\end{eqnarray}
For getting enough e-folding, e.g. ${\cal N}_{e}\sim 60$, we require inflaton period $\lambda_{\rm inf}$
\begin{eqnarray}
{\lambda}_{\rm inf}={\cal N}_{e} {1\over \sqrt{q_{+(1)}}}~.
\end{eqnarray}
After this period, we assume that $w_1$ drastically varies its value and the relevant term in \eqref{LateTime} becomes subdominant. When the term becomes comparable to other terms, the inflation stops.

\begin{figure}[t]
\begin{center}
 \includegraphics[width=.45\linewidth]{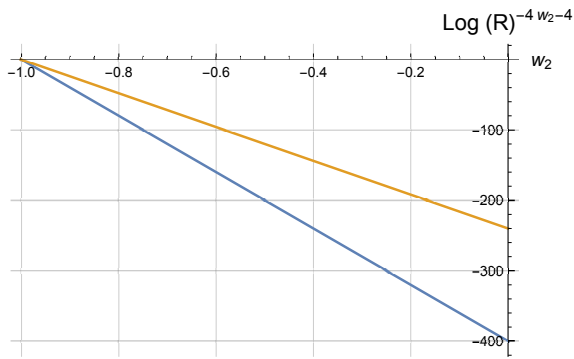}
\vspace{-.1cm}
\caption{\sl The blue and orange lines  correspond to $R=e^{100}$ and $e^{60}$. Near the region where $w_2$ is very close to zero, the functions $1/R^{4w_2+4}$ becomes drastically large. }
\label{FigLateTimew2}
\end{center}
\end{figure}

After the inflation $R$ is exponentially large, thus the most of the terms in \eqref{LateTime} becomes irrelevant. This is nothing but the wash-out by inflation. However, since we assume the thawing type behavior for the second quintessence, $w_2$ eventually approaches to $-1$. In this case, the following terms get back to relevant
\begin{eqnarray}
{\dot{{R}}^2\over R^2}=  {q_{+(2)}\over {R}^{4w_2+4}}+{l^2 \over 4} \Big(  {\Delta q_{(2)} \over {R}^{4w_2+4} }\Big)^2 +\cdots ~.
\end{eqnarray}
As one can see in the figure \ref{FigLateTimew2}, a tiny value of function for general choice of $w_2$ drastically is enlarged as $w_2$ approaches to $-1$. For example, the very small number $e^{-70}$ becomes large enough by their small powers, $[e^{-70}]^{1\over 10}\simeq 9.1\times 10^{-4}$, $[e^{-70}]^{1\over 100}\simeq 0.49$. This fact can be used to explain the smallness of the dark energy.

Finally, let us determine the quantity of the thawing quintessence in terms of data of present age as a boundary condition. We refer to the scale factor and $w_2$ at the present age as $R_0$ and $w_0$. For the quintessence to explain the small present dark energy, $q_{+(2)}$ should be
\begin{eqnarray}
{\Lambda^{(4)}_0 \over 3}\simeq  {q_{+(2)}  \over {R}_0^{4w_0+4}}~.
\end{eqnarray}

\section{Conclusions and discussions}

In this paper, we studied catalytic effects induced by quintessence in five dimensions. We computed the decay rate of metastable vacuum by using Coleman's method and the technique to treat a singular bounce solution developed by \cite{Gregory14}. We found that the decay rate is highly enhanced by the catalysis and the lifetime becomes much shorter. Since the lifetime varies from bubble to bubble, this can be seen as a dynamical selection of four dimensional expanding universe. As in \cite{Dan}, we considered on the critical bubble where the cosmological constant on the bubble is vanishing by tuning the parameters in five dimensions. This fine-tuning is a price we have to pay to get small dark energy. We also imposed the trans-Planckian censorship conjecture \cite{TCC} on the decay process. The created bubble expands eternally, it eventually violates the condition, even if the cosmological constant is small. This contradicts with quantum gravity theory and can never occur in a consistent low energy theories. Moreover, since there is no solution corresponding to four dimensional AdS spacetimes, we conclude that the only allowed bubble in a consistent theory of gravity has to have the vanishing cosmological constant. 

In light of this understanding, we studied an application of this model to incorporate the inflation mechanism and the dark energy to the four dimensional bubble universe. We introduced two types of quintessence fields, one is the thawing type which played a role of inflaton and the other is the freezing type which was used to explain the dark energy at the present age. Initial inflation driven by the thawing-type field washes out the universe, however, we claimed that the freezing-type of quintessence can contribute at late stage of the universe since it can becomes $w_2\simeq -1$ and very small number gets back to large. This idea may give us one of the possibility to explain the smallness of the dark energy. 

In our setup, there is no contribution of matter and radiation at the late stage. To engineer them, we have to incorporate the reheating process in this scenario. It would be interesting to study the gravitational reheating process\footnote{We thank Soichiro Hashiba for explaining us on this mechanism.} in the present quintessential inflation \cite{QuintessentialInflation,Ford} and produces matter and radiation in this context, and study observational consequences along the lines of \cite{Hashiba}. Since we naively assumed the time dependence of $w_{1,2}$ as shown in figure $1$. Next step we should do is to engineer an explicit model of quintessence in five dimensions and reproduce $w_{1,2}$ dependences. Clearly, this is beyond the scope of our paper, we would like to revisit this issue in separate publication.

\section*{Acknowledgments}

The authors would like to thank Norihiro Tanahashi for useful discussions and Soichiro Hashiba for explaining us quintessence inflation and related his paper. This work is supported by Grant-in-Aid for Scientific Research from the Ministry of Education, Culture, Sports, Science and Technology, Japan (JP20K03932 and JP18H01214). The authors thank the Yukawa Institute for Theoretical Physics at Kyoto University,  where the final stage of this work was done during the YITP-W-20-08 on ``Progress in Particle Physics 2020''.


\setcounter{equation}{0}
\renewcommand{\theequation}{A.\arabic{equation}}
\appendix

\section{Coleman-de Luccia bounce action in five dimensions}

\begin{figure}[t]
\begin{center}
 \includegraphics[width=.45\linewidth]{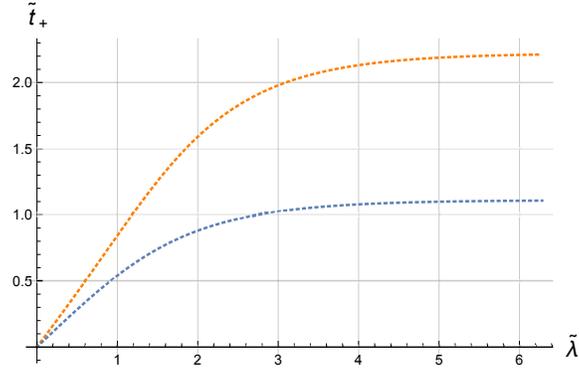}
\vspace{-.0cm}
\caption{\sl The relation between the time $\tilde{t}_+$ in AdS spacetime and the proper time $\tilde{\lambda}$ on the bubble.  The blue and orange curves correspond to parameters $\kappa_+=-2$ and $\kappa_+=-1/2$ respectively.}
\label{FigTimeRelation}
\end{center}
\end{figure}

In this Appendix, we will review the Coleman-de Luccia (CDL) bounce action \cite{CDL,KO1} by focusing on the  five dimensions. The final expression will be used  in the main text. We will adopt the same notations as the main text \eqref{normalization} and describe the position of the CDL bubble as $\tilde{r}=\widetilde{R}(\tilde{\lambda})$. $\tilde{\lambda}$ is the proper time on the bubble. For the tunneling between two vacua without quintessence, the equation of motion for the bubble reduces to 
\begin{equation}
\Big( {d\widetilde{R} \over d \tilde{\lambda} }\Big)^2=1-\widetilde{R}^2~. \label{5DCLDRdot2}
\end{equation}
The equation can be easily solved and the solution is given by $\widetilde{R}=\cos \tilde{\lambda }$. The parameter range of the proper time is $-\pi/2\le \tilde{\lambda}\le {\pi /2}$. By plugging back into the Euclidean action, we obtain the Coleman-de Luccia action
\begin{eqnarray}
  B_{CDL}=\frac{\pi}{2 G_5} \left({\gamma \over \alpha} \right)^{3} \int^0_{-{\pi\over 2}}  d\tilde{\lambda} \widetilde{R}^{2} \left(\dot{\widetilde{\tau}}_+-\dot{ \widetilde{\tau}}_-\right)\, ,\label{5CDLBounce}
  \end{eqnarray}
where relations between the proper time and times of inside and outside geometries are described by 
\begin{equation}
\dot{\widetilde{\tau}}_{\pm}=\frac{1}{ f_{\pm}}\sqrt{f_{\pm}-\Big( {d\widetilde{R} \over d \tilde{\lambda} }\Big)^2} =\frac{1}{ f_{\pm}}\sqrt{f_{\pm}-(1-\widetilde{R}^2)} \, .\label{5Dtau}
\end{equation}
In this case, $f_\pm$ is simply represented as
\begin{equation}
f_{\pm}(R)=1-\kappa_{\pm} \widetilde{R}^2~,
\end{equation}
where $\kappa_\pm$, which is negative value for an AdS spacetime, is defined by 
\begin{equation}
\kappa_+=\left( \frac{\gamma }{l \alpha}\right)^2+\frac{ (\alpha^2-1)}{ \alpha^2}  \  , \quad \ \kappa_-={\alpha^2-1\over \alpha^2}\, .
\end{equation}
Substituting this into \eqref{5Dtau}, we obtain
\begin{equation}
\dot{\widetilde{\tau}}_{\pm}=\frac{\widetilde{R}}{ 1-\kappa_{\pm}\widetilde{R}^2 } \sqrt{1-\kappa_{\pm}}\, .
\end{equation}
Solving this equation, we find the relation between time in AdS and the proper time on the bubble. The result is 
\begin{eqnarray}
t_\pm = {1\over \sqrt{|\kappa_\pm}|}\tan^{-1}\Big(  {\sqrt{|\kappa_\pm|} \over \sqrt{1+|\kappa_\pm|} }\sinh \lambda \Big)~,
\end{eqnarray}
where we used the Minkowski time.  As one can see from the figure \ref{FigTimeRelation}, even in the short period of the AdS spacetime, it corresponds to long enough time on the bubble. Therefore, when we apply the trans-Planckian censorship conjecture \cite{TCC} to the four dimensional universe on the bubble, we have to use this proper time.

Finally, we show the bounce action for the five dimensions in terms of an analytic function. We assume the decay of AdS (or Minkowski) to AdS vacua, namely, $\kappa_\pm \le 0$. In this case, the bounce action can be described in a relatively simple form:  
\begin{eqnarray}
B_{CDL}=   \frac{\pi \gamma^{3}}{2 G_5 \alpha^{3}} \left[\sqrt{1-\kappa_+}H(\kappa_+)-\sqrt{1-\kappa_-}H(\kappa_-) \right]~,
  \end{eqnarray}
where we defined 
\begin{equation}
H(\kappa)\equiv \int_0^1 dx {x^{3} \over (1-\kappa x^2)\sqrt{1-x^2}}= {1\over |\kappa|}-{1\over \sqrt{|\kappa|^3 (1+|\kappa|)}}\sinh^{-1} \sqrt{|\kappa|} ~. \nonu
\end{equation}

%
%

\end{document}